\documentclass{aastex}

\usepackage[]{natbib}

\slugcomment{Accepted by {\it The Astrophysical Journal}}

\shorttitle{WR 147}
\shortauthors{Reimer et al.}

\begin{document}

\title{Parameter constraints for high-energy models of colliding winds of massive stars: the case WR 147}

\author{A. Reimer\altaffilmark{1} \and O. Reimer\altaffilmark{1} 
}

\altaffiltext{1}{W.W. Hansen Experimental Physics Laboratory and Kavli Institute of Particle Astrophysics \& Cosmology, 
Stanford University,  452 Lomita Mall, Stanford, CA 94305-4085, USA; afr@stanford.edu, olr@stanford.edu}

\begin{abstract}

We explore the ability of high energy observations to constrain orbital parameters of long period massive binary systems
by means of an inverse Compton model acting in colliding wind environments.
This is particular relevant for (very) long period binaries where orbital parameters are often poorly known from
conventional methods, as is the case e.g. for the Wolf-Rayet (WR) star binary system WR~147 
where INTEGRAL and MAGIC upper limits on the high-energy emission have recently been presented.
We conduct a parameter study of the set of free quantities describing the yet vaguely 
constrained geometry and respective effects on the non-thermal high-energy radiation from WR~147.
The results are confronted with the recently obtained high-energy observations and with sensitivities
of contemporaneous high-energy instruments like Fermi-LAT. For binaries with sufficient long periods, like WR~147,
gamma-ray attenuation is unlikely to cause any distinctive features in the high-energy 
spectrum. This leaves the anisotropic inverse Compton scattering as the only process that reacts sensitively on the
line-of-sight angle with respect to the orbital plane, and therefore allows the deduction of system parameters even
from observations not covering a substantial part of the orbit.

Provided that particle acceleration acts sufficiently effectively to allow the production 
of GeV photons through inverse Compton scattering, our analysis indicates a preference for 
WR~147 to possess a large 
inclination angle. Otherwise, for low inclination angles, electron acceleration is constrained to be less efficient as 
anticipated here. 

\end{abstract}


\section{Introduction}
Massive early type stars (O-, early B-, Wolf-Rayet (WR) stars) are hot
($T_{\rm eff}>10000$~K), and possess some of the strongest sustained
winds among galactic objects. Among the massive early-type stars, WR-stars 
have the highest mass loss rate $\dot M\sim 10^{-4...-5}~M_\sun/$yr.
Winds of massive stars are generally supersonic, reaching terminal velocities of 
typically $v_\infty >1000-5000$~km/s \citep{Cassinelli}.
Their kinetic energy, $L_w=\dot M v_\infty^2/2$, 
however, rarely exceed 1\% of the
bolometric radiative energy output of typically $\sim 10^{38}$~erg/s.
When two such stars form a system, their winds collide, creating an interaction region 
of high temperature ($T\sim10^{7\ldots 8}$K) shocked gas. The position of this collision region
is determined by ram pressure balance, and two (reverse) shocks follow where Fermi acceleration
has been suggested \cite[e.g.,][]{Eichler93}.

Convincing observational evidence for particle acceleration to 
relativistic energies mediated by their supersonic 
winds comes from the detection of non-thermal radio emission 
\cite[e.g.,][]{Abbott86}.
This component has been interpreted as synchrotron emission on the basis of the
measured spectra (softer than the canonical value $\alpha_r\sim 0.6$,
$F_\nu\propto \nu^{\alpha_r}$) and high 
brightness temperatures of $\sim 10^{6-7}$K, and serves also as a proof for the existence of 
magnetic fields in the shock environment.
These findings triggered the development of theoretical models of non-thermal photon production
in the shock region which is exposed to  
a strong radiation field in the UV range and high wind particle densities
from the participating hot stars, and their magnetic fields. In these scenarios non-thermal 
high-energy emission from the shock region, extending well into the $\gamma$-ray domain, has been predicted
\cite[e.g.,][]{first,Eichler93,corr_analysis,bene2003,Muecke,Reimer06,Pittard06b}.

Generally, $\gamma$-rays are produced either by inelastic nucleon-nucleon interactions if
the cosmic ray nucleons reach sufficient high energies to overcome the threshold of this process,
or are produced through leptonic processes like inverse Compton scattering and non-thermal bremsstrahlung.
Typically, inverse Compton produced fluxes are found to dominate over bremsstrahlung produced ones here.
For close binary systems it has also been proposed that nucleons are accelerated in the shock up to 
EeV energies, with subsequent photodissociation reactions and/or photomeson production \cite[e.g.,][]{Bednarek05} 
that may initiate pair cascades, eventually leading to detectable $\gamma$-ray flux levels.

In the course of the last decade, several high energy phenomena have been observed supporting the occurrence of 
$\gamma$-ray production in massive star systems: the EGRET detections of $\gamma$-ray sources that are, 
although presently unidentified, found in intriguing positional coincidence with massive star binary systems
\citep{Kaul97,corr_analysis}, binary systems containing a massive star and a compact companion have been discovered
as sources of TeV-emission \citep[e.g.,][]{psr1259,LS5039,LSI,CygX1}, the southern massive star cluster Westerlund 2 
shows extended TeV-emission \cite{Westerlund2}, as does Cyg OB2 \cite{HEGRA,CygOB2}, with winds of its massive stars likely involved in powering the emission.
Recently, the very-long period binary system \object{WR 147}, consisting of a WN8(h) and a B0.5 V star, and with 
$\sim 650$pc \cite{Morris00} the second closest of all Wolf-Rayet binaries, has been observed with the MAGIC telescope for 
30.3 hours of good data. From these observations the MAGIC collaboration reported upper limits of 
$F_{>80{\rm GeV}} < 1.1\cdot 10^{-11}$~cm$^{-2}$ s$^{-1}$, $F_{>330{\rm GeV}} < 3.1\cdot 10^{-12}$~cm$^{-2}$ s$^{-1}$ and
$F_{>600{\rm GeV}} < 7.3\cdot 10^{-13}$~cm$^{-2}$ s$^{-1}$\cite{MAGIC_UL}.
Because of WR~147's proximity this system belongs to the few where high-resolution radio
observations lead to resolving the collision region between the main sequence
stars as an extended, slightly elongated non-thermal feature
on VLA and MERLIN images \cite{Dougherty96,Contreras97,Dougherty00}
in addition to the free-free emission from the spherical wind of the stars.
Furthermore, X-ray imaging with Chandra revealed an extended region at the location of ram pressure balance, and was
interpreted as the wind-wind collision zone \cite{Pittard02}.
However, despite a multitude of observations dedicated towards this object, as well as its relatively short distance to Earth,
the geometry of this system is not well known. The unknowns start with WR~147's period, being at least
1350 years \cite{setia}, the lack of clues about its argument of periastron, and the yet pending observational hints towards a determination of its orbital 
eccentricity. Proposed inclination angles, mainly deduced from modeling of the extended radio and X-ray images, cover a wide range: from
``low'' \cite[$\leq 30\degr$:][]{Dougherty03,Pittard06}  to ``large'' \cite[$> 40\degr$:][]{Williams97,Contreras99,Pittard02} with respect 
to the plane of the sky.
Consequently, any predictions for the $\gamma$-ray emission from this source must be concurrently evaluated in conjunction with
the assumed parameter space when confronting with observational high energy measurements. 

The aim of this work is to scientifically explore and relate the sensitivity of $\gamma$-ray detections or limits by operating 
(H.E.S.S./VERITAS, MAGIC, AGILE, Fermi Gamma-ray Space Telescope) instruments to the geometry (inclination, eccentricity, orbital phase) 
of the massive binary system WR~147. For this purpose we are using the emission model for colliding winds of long-period massive stars of 
Reimer et al. (2006), which self-consistently calculates the non-thermal particle
population and its radiation after diffusive acceleration of particles taken from the pool of thermal wind material.
This model includes all relevant radiative processes/losses, expansion losses and propagation effects such as 
particle convection out of the wind's
collision zone for a generally not too optically thick environment, which is shown to be appropriate for WR~147. 
As indicated in Reimer et al. (2006) inverse Compton (IC) emission well dominates over all competing $\gamma$-ray production 
mechanisms in wide massive binary systems, thereby determining their high-energy appearance. Hence we restrict our calculations to this process only. 
Both Klein-Nishina effects as well as the angle-dependence of the cross section
are taken into account, with the latter playing a decisive role for the results of our exploration.

In Sect.~2 of this paper we provide an overview of the system parameters and measurements of the non-thermal emission 
component of WR~147. A brief reminder about the main characteristics of the non-thermal emission model of Reimer et al. (2006)
for the collision region of long-period massive binary star systems is given in Sect.~3. Sect.~4 is devoted to
a parameter study of the set of free geometry describing quantities and their effects on the predicted non-thermal high-energy
radiation from WR~147. These are confronted with all information gathered so far
from measurements of WR~147's non-thermal emission component.
We conclude in Sect.~5 with our main results of this work, the constraint on the geometry of the orbit on WR~147.


\section{The knowns (and unknowns) of WR~147}

Located somewhat off the galactic plane and in spatial coincidence with the Cygnus OB2 region, although not with the nearby unidentified TeV source TeV 2032+41 \cite{HEGRA}, \object{WR 147} is 
one of the radio-brightest massive star systems known, and, after $\gamma^2$ Velorum, 
the second closest WR-star to Earth. 
Distance estimates range from $(630\pm70)$~pc \cite{Churchwell92} to $(650\pm 130)$~pc
\cite{Morris00}. We will use 650~pc as WR~147's distance in the following.
Although the binary nature of this system was already proposed in 1990 \cite{Williams90}, it was not generally accepted 
until high-resolution images 
undoubtedly established the existence of a weak Wolf-Rayet companion of type WN8(h) to the main sequence star B0.5 V.
In particular, WR~147 is meanwhile resolved into its components in the radio band by MERLIN (1.6~GHz, 5~GHz) 
\cite{Williams97}, 
VLA \cite{Churchwell92,Skinner99,Contreras99}, UKIRT K-band and HST-WFPC2 \cite{Niemela98} optical imaging. 
Notable is the dominance of the non-thermal component, interpreted as synchrotron radiation, 
between the thermal components of the stars, which points towards a colliding wind scenario operating in this system.
High-resolution X-ray
observations \cite{Pittard02} revealed an extended emission region with its centroid located in 
agreement with the location of the bow shock between the stars. The binary separation at 5~GHz was determined 
to $643\pm157$~mas, corresponding to $417~\rm{AU}/\cos{i}$ at a distance of 650~pc, where $i$ is the inclination of 
the system. 
($i=90\degr$ corresponds to the observer being located in the plane of the stars.) Here, the WR-star is separated
from the non-thermal source by $575\pm15$~mas, while the B-star's projected distance to the synchrotron source
is $68$~mas. This is found in good agreement with a wind-momentum ratio of 
$\eta = (\dot M v_\infty)_{\rm OB}/(\dot M v_\infty)_{\rm WN} \approx 0.011\pm 0.016$ \cite{Williams97} to 
$\eta = 0.02$ \cite{Pittard02}. 
From luminosity \cite[$L_{\rm bol} = 5\cdot 10^4 L_{\sun}$:][]{Morris00} and spectral 
\cite[effective temperature $T_{\rm eff}=28500$~K:][]{Crowther97} measurements 
the spectral type of the main sequence star has been estimated to
B0.5 \cite{Williams97}, and more recently proposed to O5-7I \cite{Lepine01} from HST STIS spectroscopy.
Mass loss rates for the WN star has been derived from the VLA measurements by Churchwell et al. (1992). This was
later improved by Morris et al. (2000), who found, by comparing his ISO images and 
optical data with non-LTE line models, wind filling factors $f$ of order 0.1. Because 
$\dot M \propto \dot M_{\rm true}/\sqrt{f}$ this leads to a lowering of the WR-star loss rate to 
$\dot M_{\rm WR} = 2.4 \cdot 10^{-5}$M$_\sun$ yr$^{-1}$ for a wind velocity of $v_{\rm WR}=950$~km/s \cite{Morris00}.
Setia Gunawan (2001) then used a wind-momentum ratio (of $\sim 0.014$) to conclude on the mass-loss of
the main sequence companian: $\dot M_{\rm OB}=4\cdot 10^{-7}M_{\sun}$yr$^{-1}$ for $v_{\rm OB}=800$~km/s.
On the basis of the binary separation WR~147 has to be classified as an extremely long period binary \cite{cat}, 
yet it's precise period is still unmeasured. A lower limit of 1350~yrs
was derived by Setia Gunawan (2001). No values for the eccentricity of WR~147 are documented in the literature so far.

Although attempts have been undertaken to estimate the inclination of this system, the implied range
of values for this quantity is broad: fitting the extended image of the non-thermal radio component
to a bow shock model gave values of $45\degr\pm15\degr$ \cite[VLA:][]{Contreras99}, $30\degr\pm 15\degr$ \cite[VLA:][]{Contreras04}, 
41\degr \cite[MERLIN:][]{Williams97} and $\leq 30\degr$ \cite[MERLIN:][]{Dougherty03}.
It was conjectured that the different antenna characteristics and the treatment of the system noise
may influence the modeling procedure noticable.
Practically, this means that the inclination of WR~147 is at least highly uncertain. Modeling the extended Chandra image required an inclination angle of $\sim 60\degr$ \cite[][]{Pittard02}.
Consequently, we will use
this quantity in this work as a free parameter, in an attempt to constrain its value from observational limits on 
the non-thermal high energy component of WR~147.

At present, only the radio band observations provided signatures of a non-thermal particle population in massive star systems,  
and data have been obtained in a number of measurements. We use the summary table from Setia Gunawan (2001)
for the purpose of this work.
After extracting the thermal component from the total measured one, one finds a power-law between
1-10 GHz  with
spectral index $\alpha_r\simeq 0.43-0.5$ \cite[$S_\nu \propto \nu^{-\alpha_r}$;][]{setia,Dougherty03}, and with a low-energy turn-over at $\sim 0.5-1$~GHz. 
A note of caution must be put here: the different instrument setups and capabilities, analysis methods, etc. lead to up 50\% flux differences at the various frequencies.
This applies in particular to the high-frequency data points at 22~GHz and 43~GHz, which additionally possess by themselves large measurement uncertainties. 
Suggestions \cite[e.g,][]{Dougherty03} for a high frequency turnover at these energies are therefore not to be considered established \cite[e.g.,][]{setia,Pittard06}.

For the present work we use the synchrotron energy spectrum $S_\nu \propto \nu^{-0.5}$ with a cutoff
well above 50 GHz. This corresponds to an $E^{-2}$ power law particle spectrum extending beyond a few $10^{3}$~MeV
for mGauss magnetic field strengths.

Flux variations in the light curve of WR~147 have been reported occasionally \cite{Churchwell92}, 
however, could not be confirmed \cite{setia,Skinner99}. Instead, any flux variability has rather been found to occur on long time scales with variability amplitudes not in excess of $\leq 15\%$ \cite{setia},
possibly indicating wind clumping.


\section{The model for non-thermal high-energy radiation in binary systems of massive stars}

The present work utilizes the colliding wind model of Reimer et al. (2006).
Here the supersonic winds of a massive main sequence star and its companion flow radially
from each star until they collide to form a region of shocked, hot gas where their ram pressure balances, 
separated by a contact discontinuity with a forward and reverse shock. It is this shock region
where particles may be accelerated out of the pool of thermal wind particles
with acceleration rate $\dot E/E = a = V_{\rm sh}^2 (c_r-1)/(3c_r \kappa_{\rm a})$ 
(with $\kappa_{\rm a}$ the corresponding diffusion coefficient, $V_{\rm sh}$ the shock velocity and 
$c_r$ the shock compression) by the Fermi mechanism \cite{Eichler93}.
Behind the shock the gas expands sideways from the wind collision region along the contact
surface with a convection velocity $V$ being a fraction \cite[set to 0.5 here; see][]{Reimer06}
of the wind velocity at the stagnation point. Both, diffusion and convection by the stellar winds 
operate in the collision region with
diffusion being dominant close to the stagnation point (named "acceleration region" in the following), 
and convection determining the particles' transport
further away from the downstream region (called the "convection region").
For simplicity the geometry of the collision region, being the site of the non-thermal radiation component,
is adopted as  cylinder-like. This is a sensible approximation specifically for
high energy particles, the focus of this work, since 
most relativistic particles will have lost already a significant amount of their original
energy when entering the convection region.

The location of the shock front with respect to the main sequence OB star, $x_{\rm OB}$, is determined by the
balance of the ram pressure of both winds that have generally reached their terminal velocities
($v_{\rm OB}$, $v_{\rm WN}$) in long-period binary systems:
$
x_{\rm OB} = \frac{\sqrt{\eta}}{1+\sqrt{\eta}}D
$
where
$
\eta =  \frac{\dot M_{\rm OB} v_{\rm OB}}{ \dot M_{\rm WN} v_{\rm WN}}
$
and $D$ the separation of the binaries, $\dot M_{\rm OB}$, $\dot M_{\rm WN}$ 
the mass loss rates of the WN- and OB-star, respectively. 

For a given stellar surface magnetic field $B_s$
the field strength $B$ in the collision region is assumed to follow
the well established magnetic rotator theory \cite{Weber67}, and adopting a
typical rotation velocity for early type stars of $\sim 0.1 v_\infty$ \cite{Conti,Penny}.
Estimates for $B_s$ range from below 100~G \cite{Mathys99} up to $10^4$~G \cite{Ignace98}.
In Reimer et al. (2006) we used $B_s=30$~G for WR~147 which leads to approximative equipartition field values at the shock location. For the present work we shall consider a sensible range for the values of $B_s$ (see Sect.~4).

\begin{figure}[t]
\resizebox{\hsize}{!}{\includegraphics{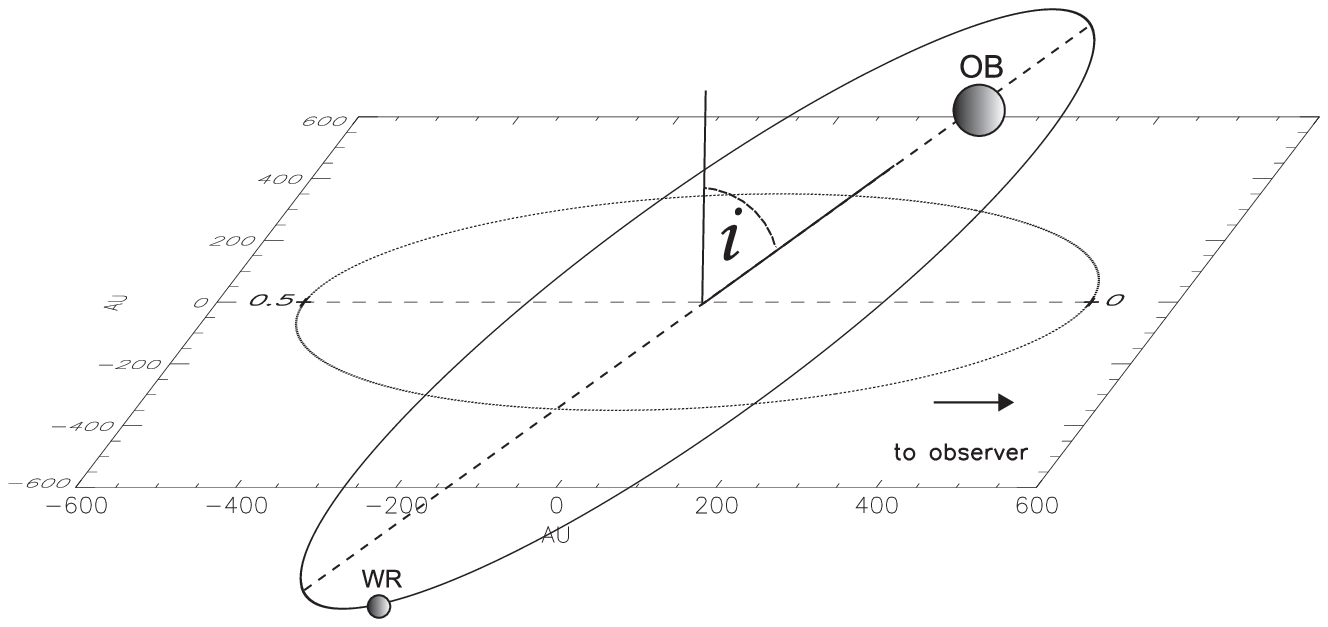}}
\caption{The binary orbit of WR 147 for an assumed inclination of $60\degr$ and eccentricity $e=0.7$.
 For details see text.}
\label{Schema}
\end{figure}

The plane of the binary system is inclined by an angle $i$ with respect
to the observer ($i=\pi/2$ corresponds to an observer in the plane of the stars; see Fig.~\ref{Schema}), and $\phi_B$, the angle
between the projected sight line and the line connecting the stars
is a measure of the orbital phase $\Phi$ of the system. Because the argument of periastron is
not known for WR~147, we will define here periastron passage at $\Phi=0=\phi_B$, and $\phi_B=0$ for the WN-star 
being in front of the B-star along the sight line.
In the co-rotating system centered on the OB-star the line-of-sight to the observer is described 
by the angles $\phi_L$ with $\tan\phi_L=\sin\phi_B\,\cot i$, and
$\theta_L$ with $\cos{\theta_L}=\cos{\phi_B}\,\sin i$.

The emitting electron spectra are calculated self-consistently by solving the steady state diffusion-loss equation 
and including wind expansion and all relevant radiative losses 
(synchrotron, IC and losses due to Coulomb scattering and relativistic bremsstrahlung) which balance the 
acceleration gains. IC losses often extend into the Klein-Nishina regime, which requires a rigorous treatment 
of this dominating loss process. In consequence, Klein-Nishina loss dominated electron spectra cutoff at higher
energies as compared to electron spectra that are treated with the Thomson approximation throughout
the whole particle energy range. 
In the convection region the particles loose energy radiatively as well as through expansion
losses while in the post-shock flow. This leads to a
dilution of the particle density and to a deficit of high energy particles in this region.
Because the stellar target photons for IC scattering arrive at the collision region from a 
prefered direction, the full angular dependence of the cross section has to be taken into account 
\cite[e.g.,][]{Reynolds82}.
The radiation yield of IC scattering depends on the scattering angle 
$\theta_{\rm ph}=\theta_{\rm sc}$ (angle between the directions of
the incoming (from the OB-star) and outgoing photons), which obviously depends on the location of the 
scattering electron as well as the orbital phase. It is
\begin{equation}
\cos\theta_{\rm ph}= 
\cos{\theta_L}\,\cos{\theta} + \sin{\theta_L}\, \sin{\theta}\, \cos(\phi-\phi_L)\, .
\end{equation}
This leads to anisotropy effects: the cutoff energy and flux variations of the IC luminosity
are dependent on the line of sight into the wind as large scattering angles tend to produce more 
energetic photons than small scattering angles. A detailed description can be found in Reimer et al. (2006).


\section{Parameter study and their constraints}

The emission model of Reimer et al. (2006) for modeling a binary system 
of known geometry (orbital phase, eccentricity $e$, and period - or binary separation, the argument of periastron, 
and the inclination angle $i$ with respect to the plane of the sky) possess five free parameters:
main sequence star surface magnetic field, diffusion coefficient $\kappa_{\rm a}$ , shock compression factor $c_r$ , 
convection velocity $V$ and normalization of the injected particle population. These are contrasted by various physical 
and observational constraints: In order to accelerate particles to the observed relativistic energies
the diffusion coefficient must be low enough to allow acceleration to overcome the Coulomb losses. 
At the same time, the Bohm-limit for the diffusion coefficient shall not be violated. 
A further observational limit may be imposed by the observed maximum photon energy of any non-thermal component.
From the highest observed frequency data point of the synchrotron component of WR~147, 43~GHz, we deduce the maximum 
electron Lorentz factor $\gamma_{e,\rm max}$ to be at least or greater than several $10^3$ for 
mGauss magnetic field strengths assuming roughly equipartition. 
To reach those energies diffusion must be at least as efficient as $\kappa_{\rm a}=8\cdot 10^{21}$, $1.4\cdot 10^{22}$, 
$2\cdot 10^{22}$, $4\cdot 10^{22}$, $1.4\cdot 10^{23}$ and $14\cdot 10^{23}$cm$^2$ s$^{-1}$
for system inclinations $i=0\degr, 30\degr, 45\degr, 60\degr, 75\degr$ and $85\degr$ ($B_s=30$G), respectively, 
and serve therefore as upper limits for $\kappa_a$.
These limits take into account the shift of the shock location away from the OB-star with growing binary separation
for increasing inclination angles ($D = D_{\rm proj}/\cos{i}$), 
and the corresponding dilution of the photon density there. The electron spectra would then cut off due to radiative losses.
The highest electron energy is attained by decreasing the diffusion time scale up to the Bohm limit.
This limit depends again on the system inclination $i$ for an increasing binary separation and shock distance to
the OB-star, and hence a decreasing field strength at the shock location, with rising $i$. The corresponding cutoff energies 
lie, within a factor 2-3, around several $10^5$~MeV ($B_s=30$~G$-100$~G). The narrowness of this range justifies the use of a fixed 
diffusion coefficient for all inclination angles without much loss of constraining power.
In the following we therefore use $\kappa_a=3.2\cdot 10^{21}$cm$^2$ s$^{-1}$ ($B_s=30$G) or 
$\kappa_a=0.9\cdot 10^{21}$cm$^2$ s$^{-1}$ ($B_s=100$G), independent of the system inclination.
We note that any value $B_s>10$~G, together with the here assumed rotator theory, is in agreement with the lower limits for the field strength derived from the Razin-Tsytovich effect.

The particle spectra normalization is limited by energy conservation and particle number conservation: 
the total energy in form of relativistic particles shall not be larger than the available total kinetic wind energy
of the system, and the number of relativistic particles injected into the system must not exceed
the number of wind particles entering the acceleration zone. The final normalization is then fixed within these limits
by the requirement to reproduce the observed synchrotron spectrum, and is therefore dependent on the chosen
magnetic field strength.

With these constraints and limits at hands we will in the following explore the effects of varying the system geometry upon 
the expected IC radiation at high energies. This will then allow us to compare our studied geometries with the 
observational data, and set constraints on the orbital parameters in the WR~147 system.

\subsection{Inclination effects on the particle spectra}

\begin{figure}[t]
\includegraphics[height=13cm]{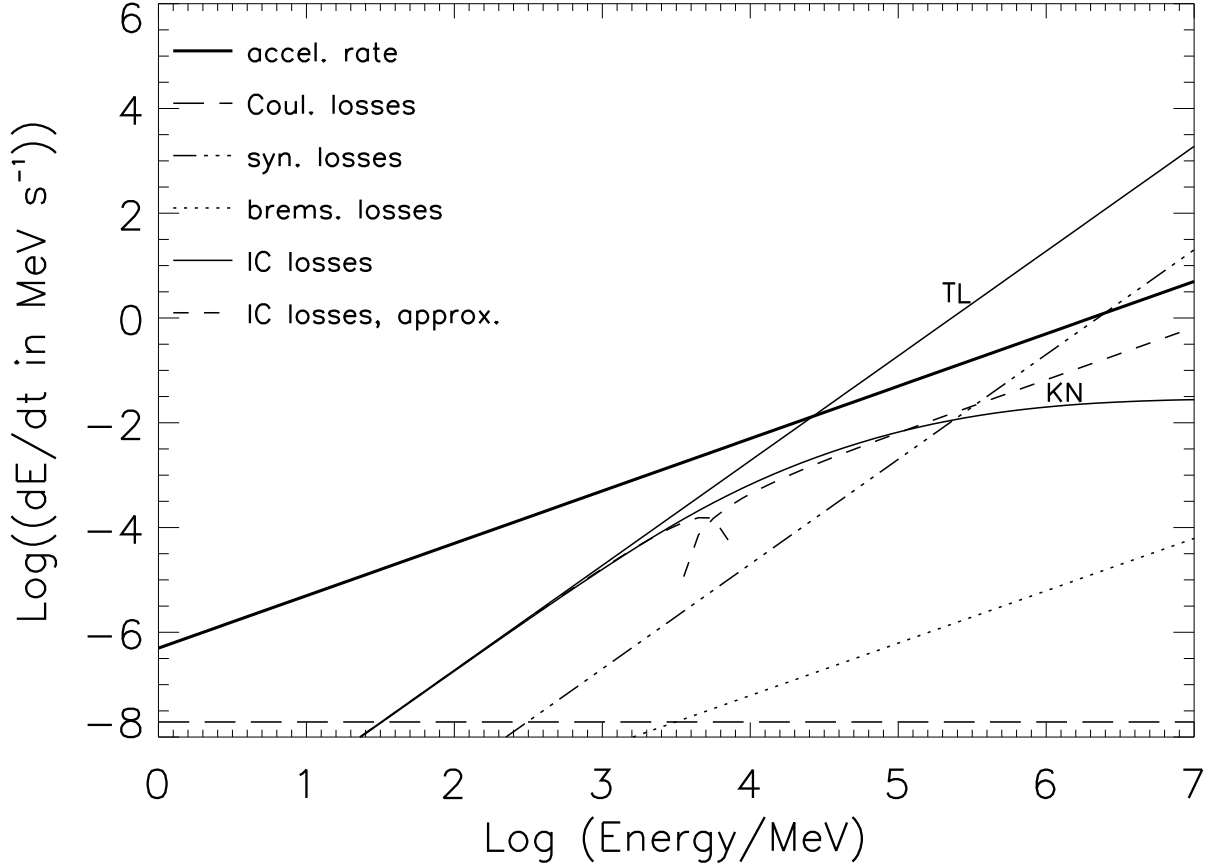}
\caption{Energy loss rates in the colliding wind region of the WR~147 binary system, assumed to be inclined by $60\degr$ with respect of the sky: IC scattering 
(thin solid lines) in the Thomson regime (TL) and
Klein-Nishina regime (KN), synchrotron radiation (dashed-triple-dotted line), relativistic electron-ion
bremsstrahlung (dashed-dotted line) and Coulomb interactions (dashed line) in comparison to the acceleration rate
(thick solid line) in the acceleration zone. The dashed line represents the IC loss 
approximation used in this work. Parameters are: $L_{\rm bol,OB}=5\cdot 10^4 L_{\sun}$,
$\epsilon_T=6.6$~eV, $\dot M_{\rm OB}=4\cdot 10^{-7} M_{\sun}$yr$^{-1}$, $\dot M_{\rm WN}=2.5\cdot 10^{-5} M_{\sun}$yr$^{-1}$,
$v_{\rm OB}=800$km/s, $v_{\rm WN}=950$~km/s, $B_s = 30$G,
$D=417{\rm AU}/\cos{i}\approx 12.5\cdot 10^{15}$cm, $x_{\rm OB}\approx 0.1D$, $B\approx 12$mG,
$n_{\rm ph,T}\approx 5.5\cdot 10^{7}$cm$^{-3}$, $N_H\approx 3.4\cdot 10^4$cm$^{-3}$,
$\kappa_a = 3.2\cdot 10^{21}$cm$^2$s$^{-1}$, $V\approx 399$~km/s, $T_0\approx 2\cdot 10^6$~s,
$r_0\approx 8\cdot 10^{13}$cm.
}
\label{acc_loss}
\end{figure}

The excellent radio images gave a projected binary separation of $\sim 417$~AU for a distance of 650~pc.
After de-projecting one obtains ``true'' star-star distances of 
$D/10^{15} = D_{\rm proj}/(10^{15}\cos{i}) = 6.25$-$71.8$cm for inclination angles
$i=0\degr-85\degr$, respectively. For this range of $i$, and together with the 
suggested wind momentum ratio 
$\eta\sim 0.014$ (see Sect.~2), the shock location $x_{OB}$ is then calculated to $x_{OB}/10^{15}$~cm$ = 0.66-7.60$
where we find magnetic field strengths of $24.9-2.2$~mG ($B_s=30$~G) and $59.1-5.1$~mG ($B_s=100$~G) for negligible field amplification (e.g. due to a run-away streaming instability in the acceleration region \cite{Bell}).
In Fig.~\ref{acc_loss} we show the importance of the various electron loss processes for a system inclination of $i=60\degr$
and $B_s=30$~G as an example. IC losses determine a broad portion of the electron spectrum, until the synchrotron losses take over
when Compton losses are already deep in the Klein-Nishina regime. At low energies (less then a few tens of MeV), Coulomb losses dominate.
The synchrotron spectra from WR~147 indicate an underlying electron spectrum with approximately $E^{-2}$. This spectral requirement
is in agreement with the assumption of a strong shock $c_r=4$, a convection velocity of $\approx 1/2 v_{\rm OB}$ and
anisotropic diffusion $\kappa_a = 4 \kappa_d$ \cite[see][for details]{Reimer06} which we shall use in the following.
Fig.~\ref{WR147_e} shows the solution of the diffusion-loss equation in the acceleration region
(of size $r_0\sim 8\times 10^{13}$~cm) 
for various inclination angles. Above $\sim 10^3-10^4$~MeV inverse Compton losses enter the Klein-Nishina regime.
The particle spectrum finally cuts off due to the particle's gyroradius exceeding the size of the acceleration region $r_0$
rather than due to radiative losses. 
(The acceleration region is treated in our model as a leaky box, with a free escape boundary at $r_0$.
$r_0$ is the distance from the stagnation point where the diffusive time scale equals the convection and therefore independent of the magnetic field.)
Since this limit scales with the magnetic field strength, the cutoff energy decreases with rising $i$ 
as a response of the higher field density at the shock location. 
In Fig.~\ref{WR147syn} we show an example
of our computed synchrotron spectrum in comparison with the data. 0.36\%-11\% ($B_s=30$G) or
0.27\%-8.3\% ($B_s=100$G) for $i=0\degr-90\degr$ of the kinetic wind energy of
the OB-star, $L_{\rm w}\approx 8\cdot 10^{35}$erg s$^{-1}$, must be used for electron acceleration
to account for the observed synchrotron flux level. Here, the extent of the emission region
was taken into account up to values of $\sim 5\times 10^{14}$~cm.

\subsection{Observing angle effects on the $\gamma$-ray spectra}

\begin{figure}[t]
\includegraphics[height=13cm]{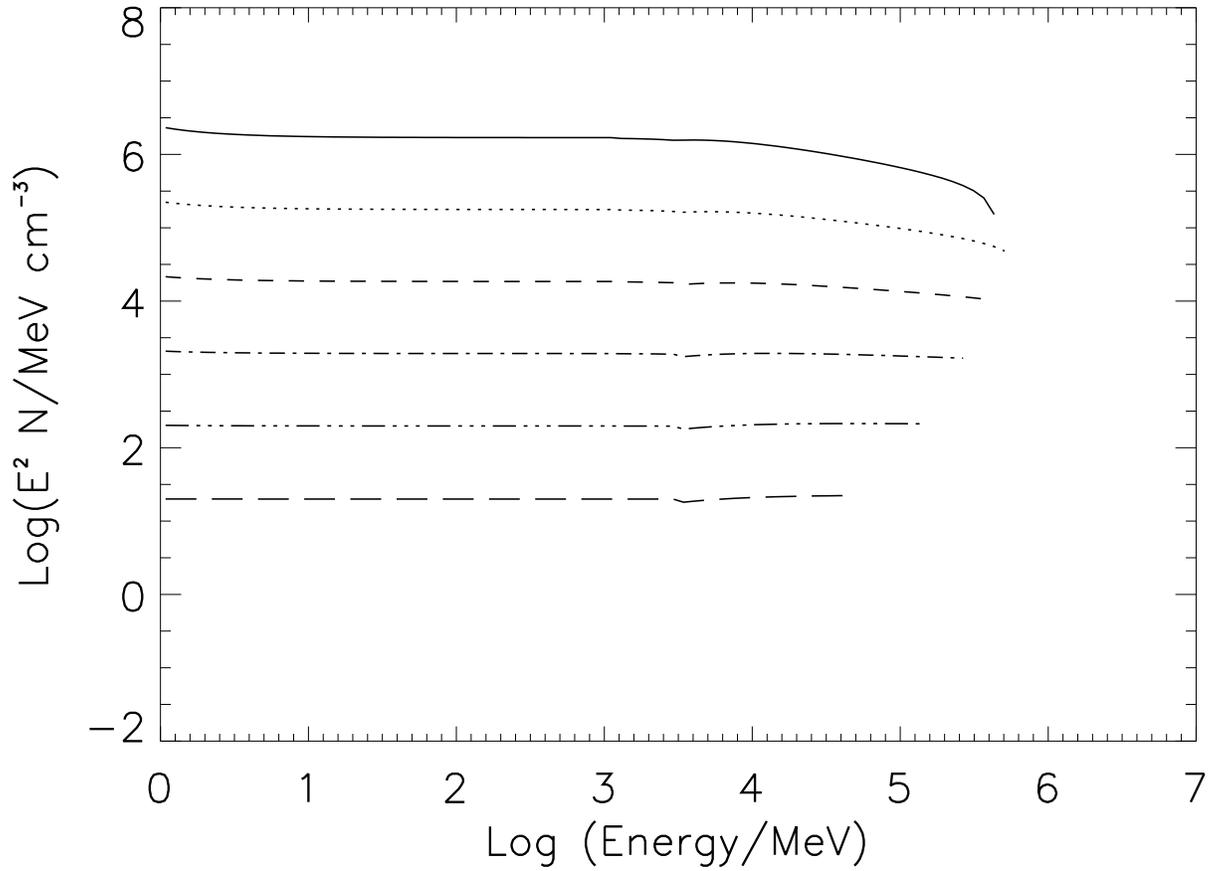}
\caption{(Non-normalized) steady-state electron spectrum for WR~147 for an inclination angle of $i=0\degr$, $30\degr$, $45\degr$, $60\degr$, $75\degr$ and $85\degr$ (upper to lower curves, each artificially separated by one order of magnitude), $Q_0=1$ and assuming a surface magnetic field for the B-star of 30~G.
See text for further parameters and discussion.
}
\label{WR147_e}
\end{figure}

\begin{figure}[t]
\includegraphics[height=13cm]{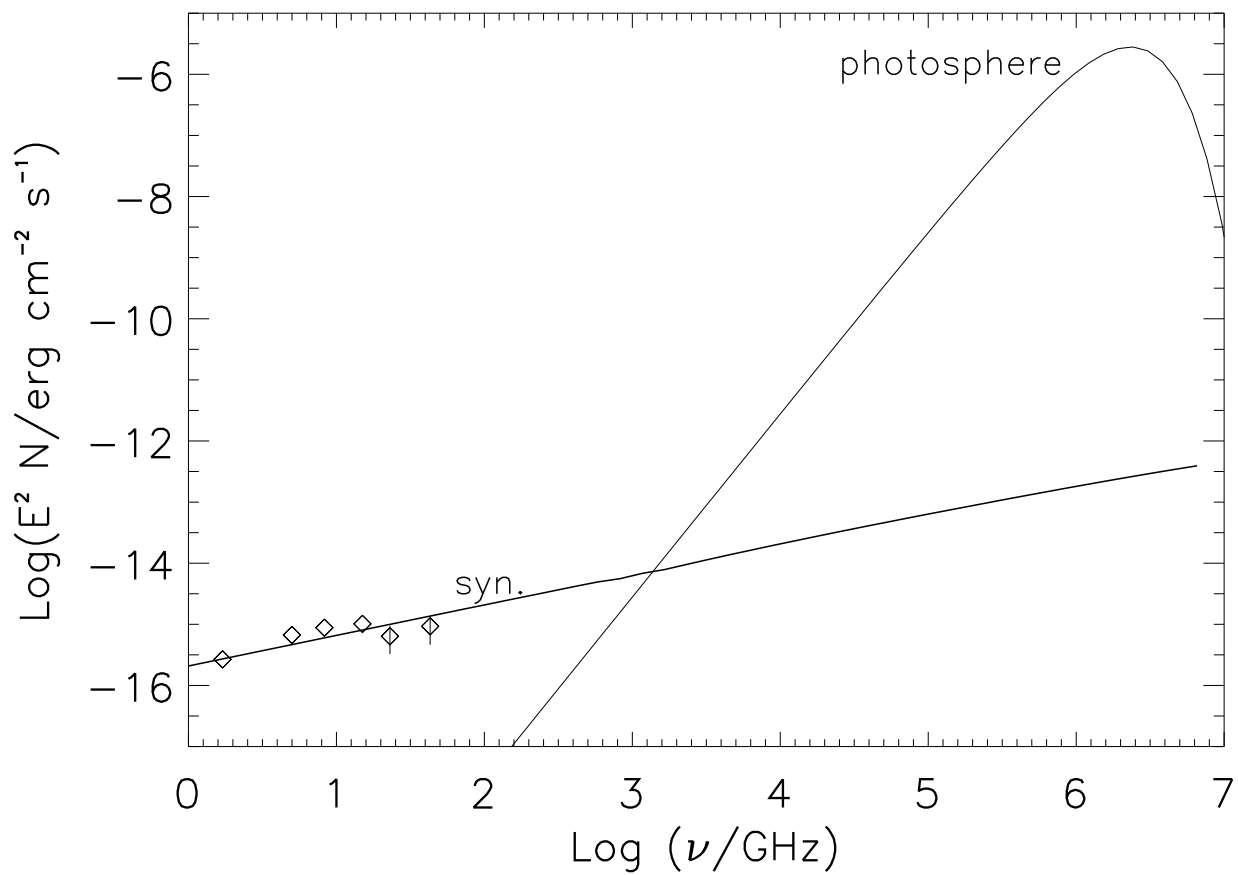}
\caption{Synchrotron spectra for WR~147 from the $i=60\degr$ electron spectrum as shown in
Fig.\ref{WR147_e} and using the $\delta$-approximation in comparison to the 
observed non-thermal radio emission taken from 
Setia Gunawan (2001a).
The bulk of the synchrotron
component is hidden below the photospheric UV radiation from
the B-star. The thermal X-ray emission from the hot shocked
gas in the collision region is not shown here.
}
\label{WR147syn}
\end{figure}

\begin{figure}[t]
\includegraphics[height=13cm]{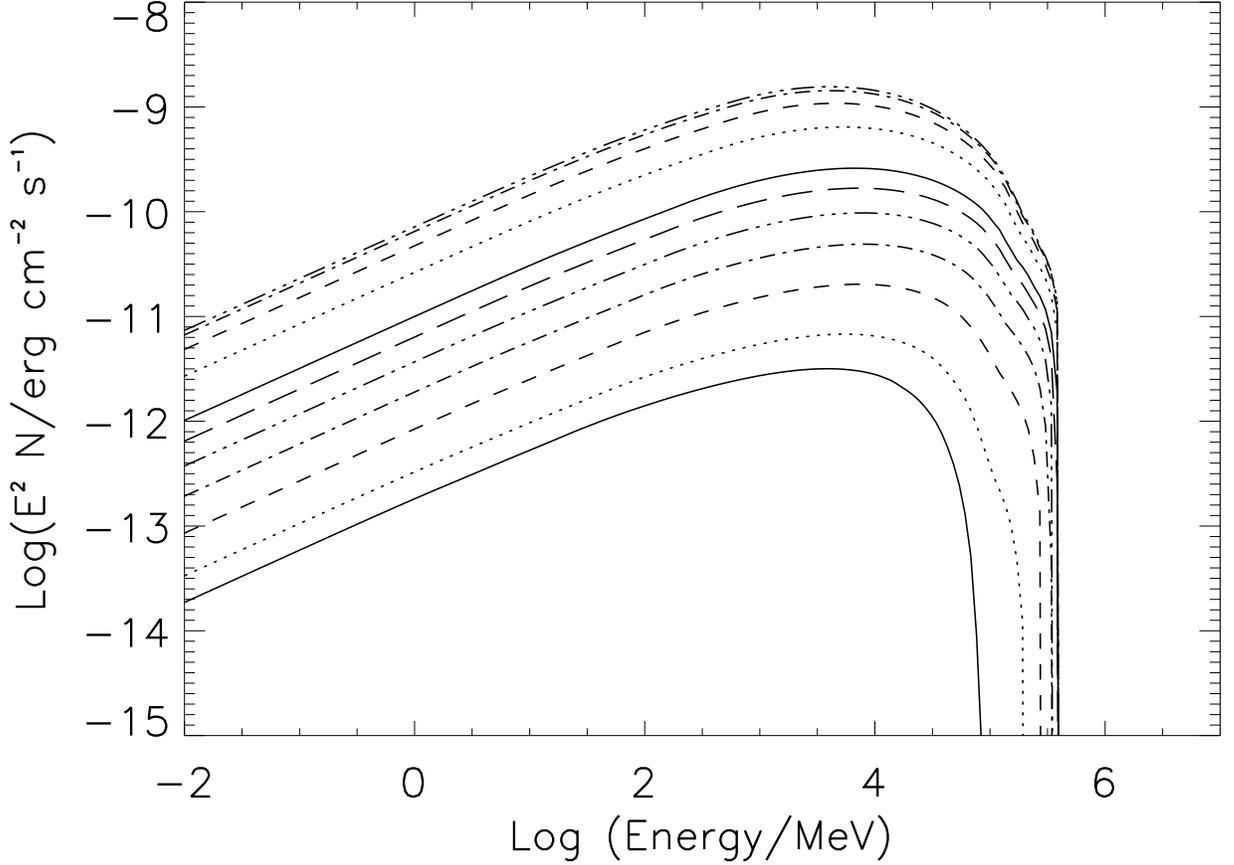}
\caption{IC spectra of WR~147 for a stellar separation of 
$6.25\cdot 10^{15}$cm 
and varying angle $\theta_L=0\degr$, $10\degr$, $20\degr$, $30\degr$, $40\degr$, $50\degr$, 
$60\degr$, $90\degr$, $120\degr$, $150\degr$, and $180\degr$ (lower to upper curves) to the observer.
0.36\% of the B-star (with an assumed surface magnetic field $B_s=30$G) wind kinetic energy has been injected 
into the acceleration region in form of electrons.
$\gamma$-ray absorption by $\gamma\gamma$ pair production is found negligible at all energies.
}
\label{WR147_IC_i00}
\end{figure}

\begin{figure}[t]
\includegraphics[height=13cm]{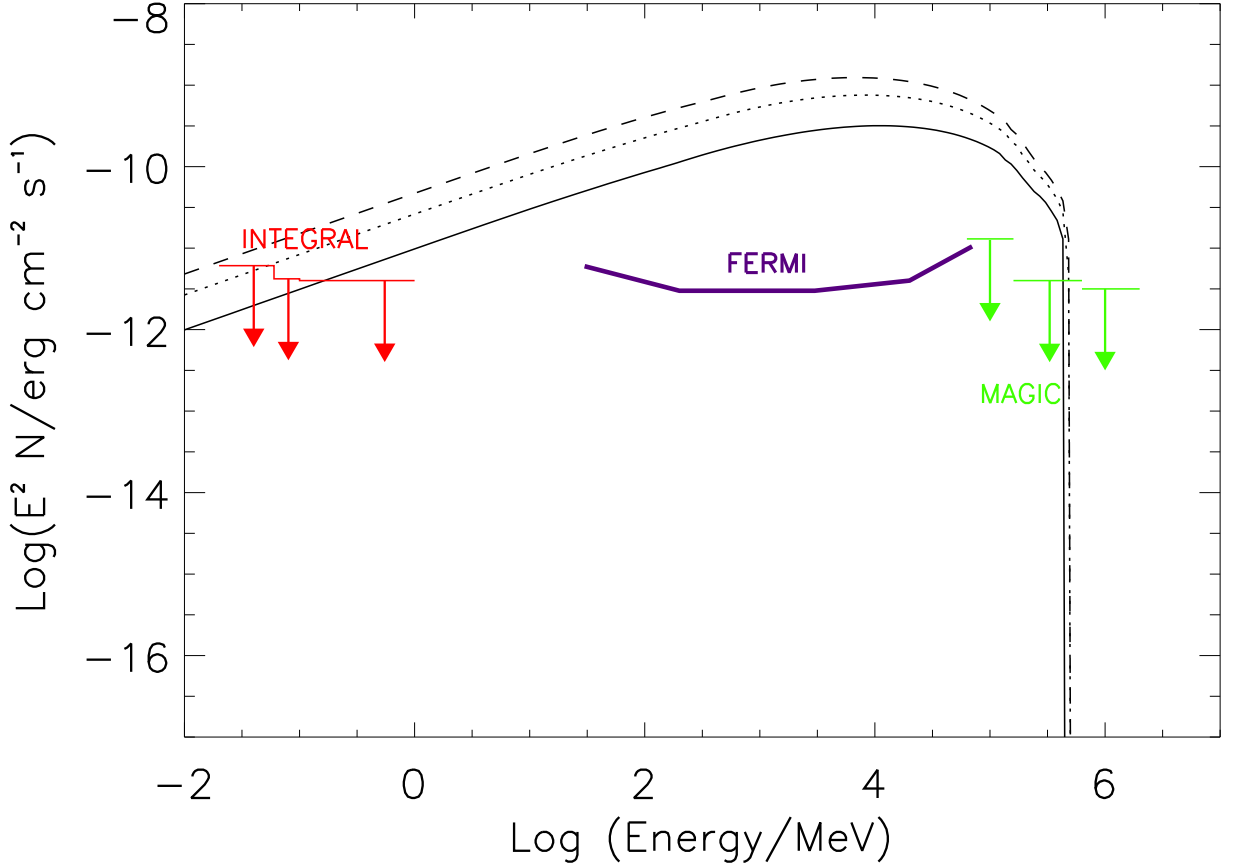}
\caption{IC spectra of WR~147 for an inclination angle of $i=30\degr$ (corresponding to a stellar separation of 
$7.22\cdot 10^{15}$cm) and geometrically possible angles $\theta_L=60\degr$, $90\degr$ and $120\degr$ (lower to upper curves) 
to the observer and compared to Fermi-LAT's sensitivity of the inner Galaxy \cite{GeVTeV}.
0.36\% of the B-star (with an assumed surface magnetic field $B_s=30$G) wind kinetic energy has been injected 
into the acceleration region in form of electrons.
$\gamma$-ray absorption by $\gamma\gamma$ pair production is found negligible at all energies.
The MAGIC (purple arrows) and INTEGRAL (red arrows) upper limits rule out this inclination angle for WR~147
for the assumed free parameter set (see text for details), or the electron spectrum at the highest energies falls off more steeply than anticipated, possibly due to}particle acceleration being not as efficient as anticipated
in this work.
\label{WR147_IC_i30}
\end{figure}

\begin{figure}[t]
\includegraphics[height=13cm]{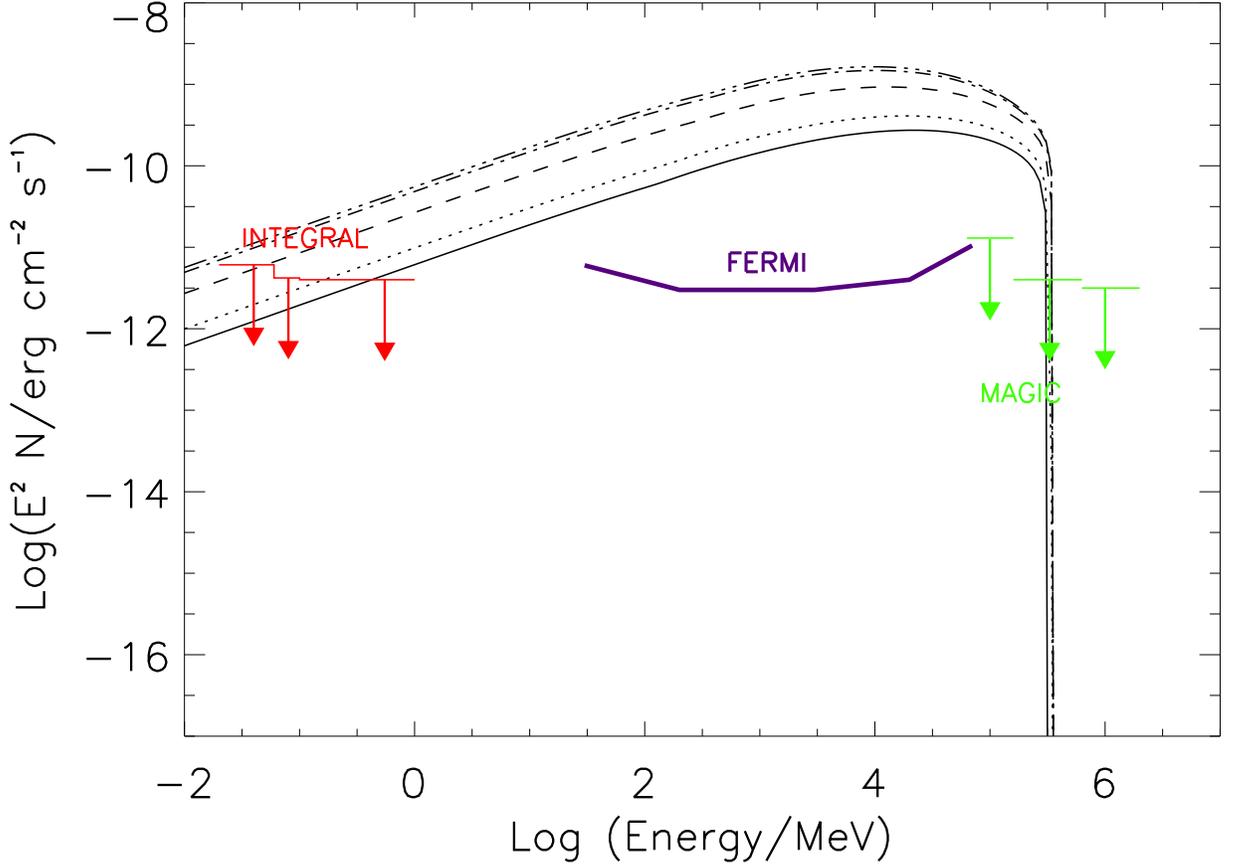}
\caption{Same as Fig.~\ref{WR147_IC_i30} but for an inclination angle of $i=45\degr$ (corresponding to a stellar separation of 
$8.85\cdot 10^{15}$cm) and geometrically possible angles $\theta_L=50\degr$, 
$60\degr$, $90\degr$, $120\degr$ and $130\degr$ (lower to upper curves).
0.55\% of the B-star wind kinetic energy has been injected 
into the acceleration region in form of electrons.
$\gamma$-ray absorption by $\gamma\gamma$ pair production is found negligible at all energies.
The MAGIC (purple arrows) and INTEGRAL (red arrows) upper limits rule out this inclination angle for WR~147
for the assumed free parameter set (see text for details), or particle acceleration is not as efficient as anticipated
in this work.
}
\label{WR147_IC_i45}
\end{figure}

\begin{figure}[t]
\includegraphics[height=13cm]{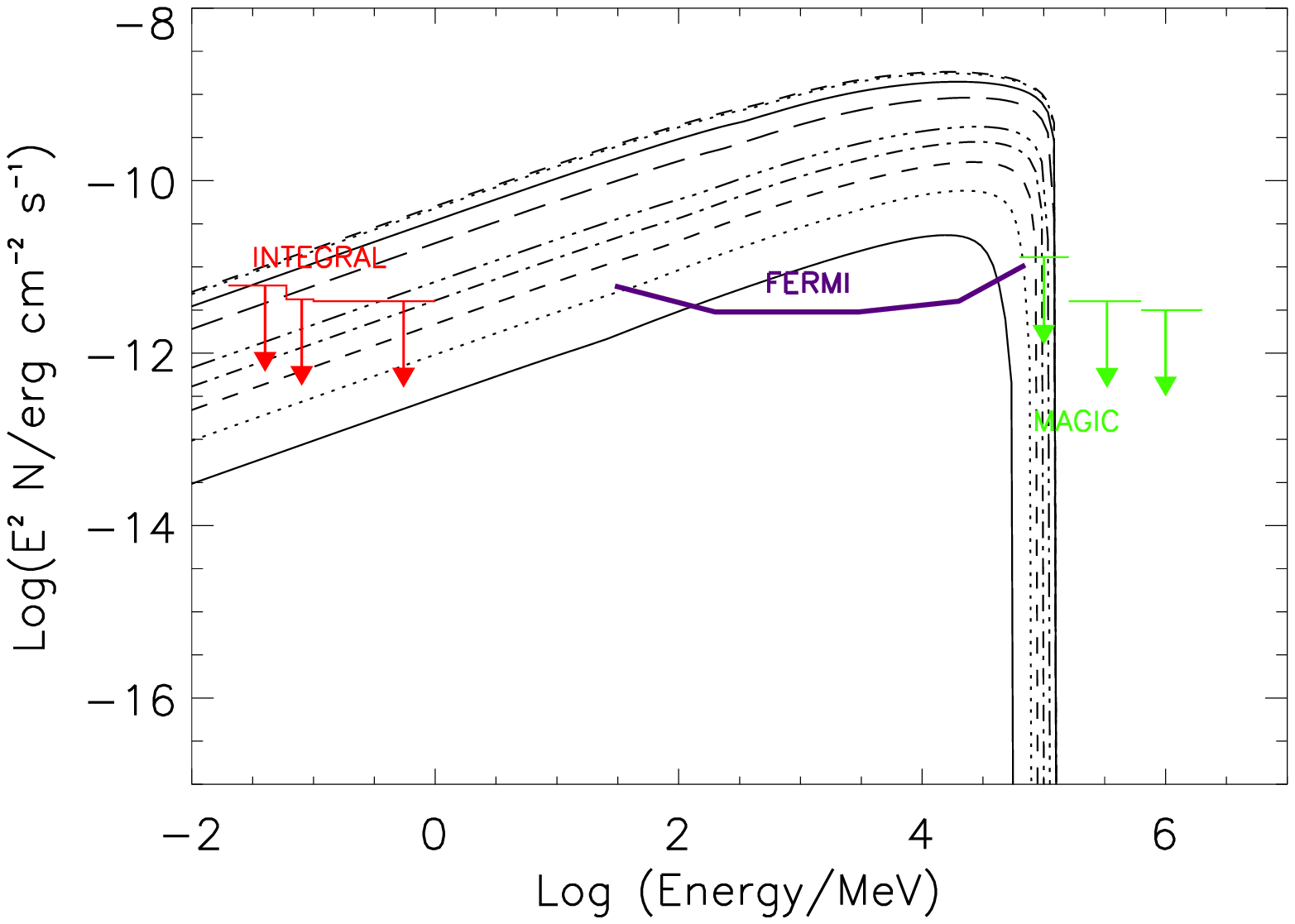}
\caption{Same as Fig.~\ref{WR147_IC_i30} but for an inclination angle of $i=75\degr$ (corresponding to a stellar separation of 
$2.42\cdot 10^{16}$cm) and geometrically possible angles $\theta_L=20\degr$, $30\degr$, $40\degr$, $50\degr$, 
$60\degr$, $90\degr$, $120\degr$, $150\degr$, and $160\degr$ (lower to upper curves).
2.23\% of the B-star wind kinetic energy has been injected 
into the acceleration region in form of electrons.
$\gamma$-ray absorption by $\gamma\gamma$ pair production is found negligible at all energies.
The MAGIC (purple arrows) and INTEGRAL (red arrows) upper limits constrain the available parameter space to
$\theta_L<50\degr$.
}
\label{WR147_IC_i75}
\end{figure}

\begin{figure}[t]
\includegraphics[height=13cm]{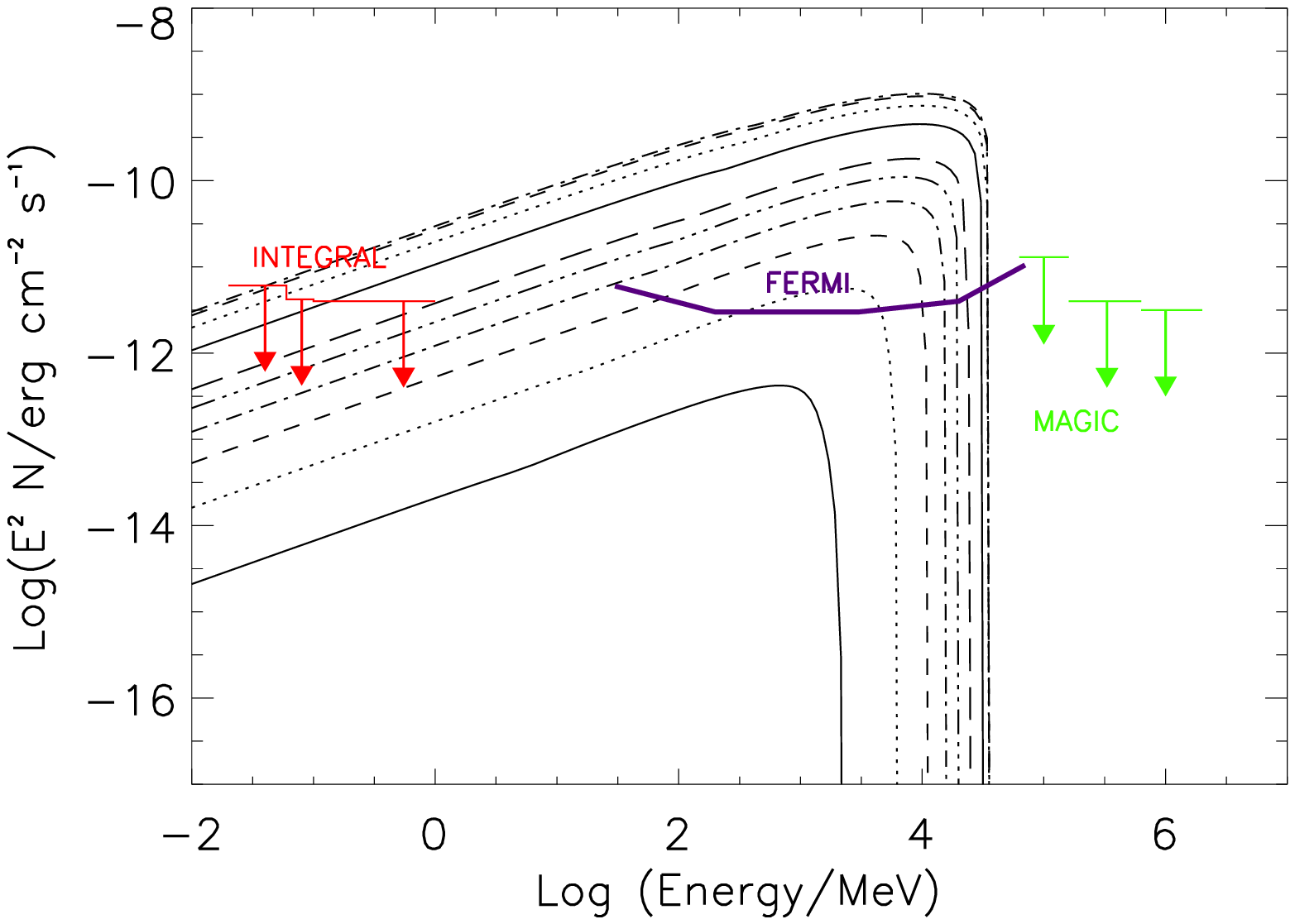}
\caption{Same as Fig.~\ref{WR147_IC_i30} but for an inclination angle of $i=85\degr$ (corresponding to a stellar separation of 
$7.18\cdot 10^{16}$cm) and geometrically possible angles $\theta_L=10\degr$, $20\degr$, $30\degr$, $40\degr$, $50\degr$, 
$60\degr$, $90\degr$, $120\degr$, $150\degr$, and $170\degr$ (lower to upper curves).
11\% of the B-star wind kinetic energy has been injected 
into the acceleration region in form of electrons.
$\gamma$-ray absorption by $\gamma\gamma$ pair production is found negligible at all energies.
The INTEGRAL (red arrows) upper limits constrain the available parameter space to
$\theta_L<60\degr$, the MAGIC upper limits does not provide any constraint.
}
\label{WR147_IC_i85}
\end{figure}

For anisotropic IC scattering, as is the case in the present systems, the resulting IC flux levels
and maximum photon energies for a given electron Lorentz factor depend strongly on the scattering angle $\theta_{\rm ph}$. Here electrons see incoming
photons from a variety of angles, which leads to raising or lowering the seed photon density accordingly 
in the electron rest frame after Lorentz transformation. 
Therefore the maximum scattered power goes into the $\theta_L=180\degr$ direction. Similarily for the
energy of the photons (in the Thomson regime): for small viewing angles the electrons are forward scattering 
the soft radiation that is less energetic in the electron rest frame than in the head-on case because of the 
Lorentz transformation. 
With Eq.~1 we find the maximum scattered power and photon energy when $\theta_L=180\degr$, i.e. the WN star being behind the OB-star 
along the sight line. At this angle the photons interact with the electrons primarily head-on, and end up therefore more energetic
after the scattering event. 
Fig.~\ref{WR147_IC_i00} demonstrates the decreasing IC flux level and maximum scattered photon energy
with declining viewing angle $\theta_L$ between the sight line and the line connecting the stars for a fixed binary separation.
From a given binary separation and the observed projected star-star distance one can infer the system inclination.
Geometrical considerations ($\cos{\theta_L}=\cos{\Phi_B}\sin{i}$) then provide a range of possible viewing angles $\theta_L$.
The viewing angle can reach higher values for large inclination systems. 
On the other hand, a large inclination increases the true binary separation,
leading to a weaker magnetic (and diluted photon field) at the shock location, and therefore to lower maximum particle energies 
(see Fig.\ref{WR147_e}) 
and Compton scattering rates. This interplay is demonstrated in  Fig.~\ref{WR147_IC_i30}-\ref{WR147_IC_i85}: the maximum 
gamma-ray energy decreases with
increasing $i$ as the maximum electron energy does (see Fig.\ref{WR147_e}). For $i=85\degr$ the electron spectrum
cuts off already at $\sim 3\cdot 10^4$~MeV, too low to generate $\sim 100$~GeV photons in this setting. 

Our model-predicted fluxes are compared to typical sensitivity limits of instruments operating in the indicated energy range, 
in addition to the recently obtained INTEGRAL and MAGIC upper limits.
For the binary separation that corresponds to $i\geq 80\degr$
the MAGIC upper limit has no constraining power on the system geometry, irrespective of orbital phase and eccentricity.
Furthermore, the overall flux level exhibits a declining trend with increasing $i$, diminishing also the 
constraints imposed by the INTEGRAL limits upon the model settings if the observer is located sufficiently 
close to the plane defined by the two stars. 
Specifically, we find the following constraints from the X-ray and $\gamma$-ray upper limits for WR~147 including the geometrical constraints:
the MAGIC upper limit allows system settings with $i\geq 80\degr$, or 
for $i\sim 70\degr-75\degr$ the allowed range $\theta_L$ increases sucessively from $\leq 20\degr$ to $\leq 50\degr$, all assuming $B_s=30$~G.
The hard X-ray limits constrain the geometry to $\theta_L\leq 40\degr$ for $i\leq70\degr$ and $\theta_L\leq 50\degr-60\degr$ for $i=75\degr-85\degr$.
The angle $\Phi_B$ which measures the orbital phase for a given system eccentricity (and hence orbital period) is determined by the knowledge of $\theta_L$ and $i$.
The corresponding orbital phases (for an argument of periastron of zero) that meet the derived ($\theta_L, i$) constraints range from $0$ to $\sim 0.4$ depending on the system eccentricity. 
Accordingly, the orbital period can be as low as $\sim 10^3$ years or as high as $\sim 10^7$ years. 
Adopting higher fields, e.g. $B_s=100$~G, the range of allowed settings increases further:  $i\geq 70\degr$, or
for $i$ around $\sim 60\degr$ the allowed $\theta_L$ ranges from $\leq 30\degr-40\degr$.
Also here the hard X-ray limits are only more constraining than the MAGIC limit for systems at large inclination angles:
if $i=75\degr-85\degr$, the viewing angle must be constrained to $\theta_L\leq 80\degr-100\degr$. The corresponding orbital phases
are mostly close to 0, but can occasionally reach values up to $\sim 0.5$, and periods are again from $\sim 10^3$years to $\sim 10^7$years,
depending on WR~147's eccentricity. Any estimate for the system eccentricity given the observational constraints set by the presented analysis is therefore coupled with the orbital phase.
 
In summary, from our analysis it appears that large inclination angles are preferred for the WR~147 system if particle acceleration
meets the exploited optimal conditions. On the other side, if WR~147 is barely inclined with respect to the plane of the sky, particle acceleration
can not be as efficient as anticipated here. 

This study demonstrates the potential of measurements within the IC component in colliding wind binary systems who show 
clear signatures of synchrotron radiation, for constraining geometrical quantities which otherwise are often difficult to gather, specifically in long period systems.
For geometrically well known systems, it offers the possibility to constrain the mostly rarely understood particle
acceleration efficiency.
Clearly, for WR~147 the Fermi LAT energy range and expected sensitivity\footnote{http://www-glast.slac.stanford.edu/software/IS/glast\_lat\_performance.htm}
over a considerably long mission timeline seems promising to ultimately put rather decisive constraints on this system's geometry.
In addition, due to its good sensitivity over several orders of magnitude of energy, Fermi LAT may probe the cutoff
energy of the putative high-energy photon component, thereby providing further constraints on the particle acceleration efficiency in 
wind-wind collision regions.

Long integration times $t_{\rm exp}$ in short period ($T$) binaries smear out any flux and spectral variations from a varying sight-line with
orbital phase, resulting in a lack of the required observables (flux/spectrum for a given phase) to pin down geometrical system parameters. 
In long period binaries with
$T\gg t_{\rm exp}$ on the other hand, significant orbital flux variations during the exposure time are not expected. Therefore long-duration system 
geometries can be deduced in this case, even if long integration times are required. If in addition the orbital period, or orbital phase at the time 
of observation, is known, constraints on the system eccentricity can be deduced.
Such observations performed at several phases and sufficient equally distributed throughout the system's orbit may intensify
the constraining power of this method.

\subsection{$\gamma$-ray opacity}

Above several tens of GeV $\gamma\gamma$-pair production can a priori not be neglected in the dense UV environment
of massive stars. Since the corresponding cross section is dependent on the interaction angle,
the expected attenuation therefore relies on the system geometry, most important, on the distance of the shock location 
with respect to the main sequence star. For WR~147, being an extremely long-period binary with a presumably small 
eccentricity, the photospheric UV radiation field at the location of the collision region is found low enough 
($\sim 10^{6-8}$ cm$^{-3}$) to allow here the neglect of $\gamma$-ray absorption due to photon-photon collisions. 
The value of $\tau_{\gamma\gamma} < 0.015$ is not exceeded in any of the considered cases, corresponding to a 
flux throughput of at least 98\% for WR~147.

\section{Summary and conclusions}

Despite its close proximity the system geometry (including inclination, period, eccentricity) of the very-long period binary WR~147 is yet not well known.
It belongs to a group of massive binary systems whose radio emission shows a non-thermal component. From radio imaging the location
of this component has been associated with the wind collision region, and the projected binary separation is well constrained.
WR~147 is one of the prime candidates for gamma ray production in this collision zone \cite{GLAST-Symp}. VHE observations with the MAGIC
telescope have been made in 2007 for 30.3 hours of good data, which resulted in upper limits at $F_{>80{\rm GeV}} < 1.1\cdot 10^{-11}$~cm$^{-2}$ s$^{-1}$, 
$F_{>330{\rm GeV}} < 3.1\cdot 10^{-12}$~cm$^{-2}$ s$^{-1}$ and
$F_{>600{\rm GeV}} < 7.3\cdot 10^{-13}$~cm$^{-2}$ s$^{-1}$\cite{MAGIC_UL}. Further observational constraints of WR~147's non-thermal component
are coming from INTEGRAL upper limits at hard X-rays and the GHz frequency radio synchrotron measurements.
In this work we explored the potential of gamma ray measurements to constrain the system geometry by means of an inverse Compton model
for high energy photon production. For this purpose we assume a high efficiency of particle acceleration that generally 
allows the production of GeV photons. Within this scenario we find:

\begin{itemize}

\item For the chosen rapidness of electron acceleration the particle spectra are cut off due to the size constraint of the acceleration region rather than due to radiative losses. In this case the cutoff energy scales with the magnetic field strength in 
the acceleration region, and we find the highest cutoff energy for low inclination systems.

\item The MAGIC upper limits at energies above 80 GeV therefore lacks constraining power for large inclinations $\geq 80\degr$ for WR~147.

\item Provided particle acceleration acts sufficiently effective to allow the production of GeV photons through inverse Compton scattering, 
then our analysis indicates a preference for WR~147 to possess a large inclination angle. However, if $i$ is small in WR~147, electron
acceleration can not be as efficient as anticipated here. 

\item Increasing the magnetic field strength in the collision region widens the range of parameter values for the WR~147 system geometry.

\item The synchrotron data from WR~147 imply the existence of $\geq 10^3$ MeV electrons, and therefore the existence
of at least hard X-ray photons produced by those particles.
Sensitive hard X-ray/soft gamma ray measurements have therefore the potential to constrain the WR~147 system geometry, 
independent of the assumed acceleration efficiency.
 
\item Gamma ray attenuation for this very-long period binary system is not anticipated to cause distinctive observational 
signatures in the high energy spectrum. Any phase-locked flux variations at $\gamma$-ray energies are therefore likely to be
attributed to the anisotropic nature of the IC process.

\end{itemize}

With a set of eight free parameters (system inclination and eccentricity, orbital phase and period, OB-star surface magnetic field, 
diffusion coefficient, shock compression factor and convection velocity) within the framework of the IC model 
for high energy photon production \cite[e.g.][]{Reimer06}, 
to be confronted with the observationally constrained quantities (normalization and spectral index of synchrotron spectrum, 
projected binary separation, MAGIC and INTEGRAL upper limits) it is clear that an in-depth determination of the so far barely 
known geometry of the WR~147 system must await further information of from measurements of the non-thermal high-energy emission component.

\acknowledgments
This work is supported by the National Aeronautics and Space Administration under contract NAS5-00147 with Stanford University. We thank Diego Torres for reading this manuscript and valuable discussions, and the referee for a constructive report.



\clearpage

\clearpage

\clearpage

\clearpage

\clearpage

\clearpage

\clearpage

\end{document}